\documentclass[10pt,journal,compsoc]{IEEEtran}
\usepackage{float}
\usepackage{listings}
\usepackage{url}
\usepackage{algorithm}
\usepackage{algorithmic}
\usepackage{booktabs}
\usepackage{hyperref}

\makeatletter
\newcommand\fs@norules{\def\@fs@cfont{\bfseries}\let\@fs@capt\floatc@ruled
  \def\@fs@pre{}%
  \def\@fs@post{}%
  \def\@fs@mid{\kern3pt}%
  \let\@fs@iftopcapt\iftrue}
\makeatother
\floatstyle{norules}
\restylefloat{algorithm}

\ifCLASSOPTIONcompsoc
  \usepackage[nocompress]{cite}
\else
  \usepackage{cite}
\fi
\usepackage{enumitem}
\ifCLASSINFOpdf
   \usepackage[pdftex]{graphicx}
\else
\fi
\begin{document}
\title{SmartSON:A Smart contract driven incentive management framework for Self-Organizing Networks}

\author{Abdullah Yousafzai, and Choong Seon Hong,~\IEEEmembership{Senior Member,~IEEE}
       
\IEEEcompsocitemizethanks{\IEEEcompsocthanksitem Abdullah Yousafzai is with the Networking Lab, Kyunghee University, Republic of Korea.\protect\\
E-mail: see dr.abdullah@khu.ac.kr
\IEEEcompsocthanksitem Choong Seon Hong is Professor at the School of Electronics and Information, Kyung Hee University, Republic of Korea.}
\thanks{Manuscript received August 26, 2020.}}

\markboth{Networking Lab, Kyung Hee University, AUG~2020}%
{Yousafzai \MakeLowercase{\textit{et al.}}: SmartSON: A Smart contract driven incentive management framework for Self-Organizing Networks}
\IEEEtitleabstractindextext{
\begin{abstract}
This article proposes a self-organizing collaborative computing network with an approach to enhance the expectation of a collaborating node for joining the self-organizing network. The proposed approach relies on Ethereum cryptocurrency and Smart Contract to enhance the expectation of collaborating nodes by monetizing the services provided to the self-organizing network. Furthermore, an escrow based smart contract is formalized in the proposed framework to sustains the monetary trust issue between collaborating nodes. The proposed scheme can enforce an autonomic incentive management mechanism to any type of self-organizing networks such as self-organizing clouds, ad-hoc networks, self-organizing federated cloud networks, self-organizing federated learning networks, and self-organizing D2D networks to name a few. Considering the distributed nature of these self-organizing networks and the Ethereum blockchain network, a distributed agent-based methodology is materialized in the proposed framework. Following this, a proof of concept implementation for the general case of a self-organizing cloud is presented. Lastly, the article provides some insights into possible future directions using the proposed framework.

\end{abstract}

}

\maketitle

\IEEEdisplaynontitleabstractindextext
\IEEEpeerreviewmaketitle

\IEEEraisesectionheading{\section{Introduction}\label{sec:introduction}}
\IEEEPARstart{T}{he} technological evolution of computer networks and computing devices has enabled a whole new range of cooperative and collaborative wired/wireless networking applications \cite{chen2019qoe,wu2019design,lee2019collaboroid,delgado2019cloud}. For instance, desktop computers integrating volunteer computing and peer-to-peer (P2P) networking into cloud architectures anticipating an architecture of a gigantic self-organizing cloud (SOC) to reap the huge potential of untapped commodity computing power over the Internet \cite{6200263}. In addition to this, we consider a federated cloud infrastructure, which makes it possible for a data center to extend its total operational capacity by subcontracting additional resources from collaborating data centers, making the infrastructure a federation of Clouds \cite{bohn2020nist}. In these referenced architectures each participant may autonomously act as both resource consumer and provider, and this stands true for ad-hoc networks also. Services built on top of a centralized architecture may suffer denial-of-service (DoS) attacks \cite{fox2009above}, unexpected outages, and limited pooling of computational resources. On the contrary, federated computing systems can easily aggregate huge potential computing power to tackle grand challenge science problems \cite{korpela2012seti}. Furthermore, with the advent of technologies such as network function virtualization (NFV), virtual network function (VNF), fog computing, and federated learning the true spirit of self-organizing ubiquitous networked applications can be anticipated. Given this, self-organizing networks (SON) connects a large number of computing environments by a P2P network, by self-organizing we assume a network that requires no or minimum human intervention for sustainability.  However, SON suffers from the problem of the motivation of the resource providers i.e. the motivation of collaborating resource provider node to provide their computing services to resource consumers.

In SON, each participating node act either as a resource provider or a resource consumer. Resource consumers operate autonomously for locating resource provider nodes offering their desired service/resource over the network to offload/delegate some of their tasks. Meanwhile, the resource consumer could utilize multiple resource instances from other resource providers if required.  In this respect, how a resource provider will be compensated/rewarded for sharing its computational resources and how resource consumers will be charged keeping in view the ad-hoc, self-organizing, and distributed nature of the environments under the probe. The accounting and billing for the reward and retribution using monetary bills and conventional banking are un-practical due to the ad-hoc, self-organizing, and distributed nature of the networks under consideration. However, the incentives to the resource provider based on its provisioned resources to the resource consumers in the network can be materialized by using smart contracts \cite{buterin2014next} and decentralized ledger-based crypto-currencies \cite{wood2014ethereum}.

The motivation factor discussed in the previous paragraph underpins the focus of this article, wherein we show how smart contracts, cryptocurrencies, and blockchain will enable incentive management services for such self-organizing and federated networks. The overall objective is to demonstrate the practicality of the proposed agent-based framework through the use of smart contracts and decentralized ledger-based crypto-currency. Smart contracts can debit and credit crypto-wallets of the resource providers and consumers based on programmed events and functions which reside in the contract. Through this method, a smart contract will be debited from the resource consumer crypto-currency wallet for the consumption of services/resources according to the valuations. Likewise, the crypto-currency wallet of a resource provider will be debited by the smart contract for the services/resources provided by the provider to the consumer in the network. More particularly through the blockchain, we ensure that the incentive management service has the attributes of consensus, provenance, ownership, immutability, finality, and access control attributes. The proposed scheme represents an important solution for improving the motivation of a resource provider node to join an ad-hoc, federated, and self-organizing environment and maintain itself in the environment and ensuring fair incentive rewards and retribution.

The remainder of this article is organized as section \ref{sec:rw} presents the related work. Section \ref{sec:pf} details the proposed framework. Section \ref{sec:poc} presents the proof of concept implementation details. Lastly, section \ref{sec:conc} concludes the article with future directions.

\section{Related Work}
\label{sec:rw}
Research on volunteer distributed computing models is multifaceted ranging from volunteer computing architectures \cite{anderson2019boinc,cunsolo2009volunteer,6200263}, resource discovery \cite{ghafarian2013cycloidgrid,lazaro2010flexible}, resource allocation \cite{rossi2019towards,xu2019dynamic}, security \cite{shota2019simulation}, incentive management \cite{yousafzai2016directory}. Edinger et. al \cite{edinger2019money} studies the effect of monetary incentives in P2P and volunteer computing models using conceptual and practical implications. Edinger et. al finds monetary incentives can enhance the intrinsic motivation to share resources when sharing takes place amongst anonymous users. However, keeping in view the distributed nature of volunteer computing blockchain-based crypto-currency \cite{nakamoto2019bitcoin} becomes a de-facto choice for the authors based on Edinger et. al hypothesis while considering the distributed and anonymous nature of Blockchain along with consensus, provenance, ownership, immutability, finality, access control attributes, and most importantly market value. 

Based on this literature review the research gap was obvious to utilize blockchain-based cryptocurrencies in volunteer computing models to enhance the motivation for resource sharing in the volunteer overlay.  In this regard, we conceptualize "SmartSON: A Smart contract-driven incentive management framework for Self-Organizing Networks". SmartSON is a generalized blockchain-powered agent-based self-organizing network model for volunteer computing.

\section{Proposed Framework}
\label{sec:pf}

\begin{figure}[H]
      \includegraphics[scale=0.5]{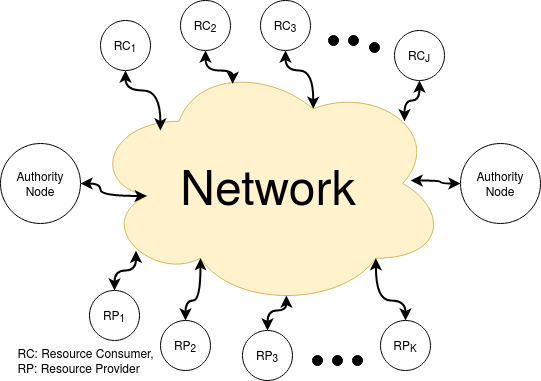}
  \caption{Abstract bird-eye view of the envisioned self-organizing network.}
  \label{figure1}
\end{figure}

We begin by describing, and giving intuition about the incentive management mechanism based on escrow, smart contract, and crypto-currencies for self-organizing networks in section \ref{sec:intution}. The rest of the sections presents the remaining three participants in the proposed framework i.e. the Authority Node, the Resource Consumers, and the Resource Providers. The smart contract resides in the ethereum crypto-currency network, while the rest are the conceptual/physical entities in the self-organizing network. These entities are sometimes referred to as nodes that are equipped with specialized software agents, these Agents follow the complete Foundation for Intelligent Physical Agents (FIPA) specification \cite{poslad2007specifying}, for distributed heterogeneous agent management and communication. These agents interact with users and with other agents available in the network to construct the self-organizing overlay network over the underlying network, enabling a distributed agent platform according to the FIPA specifications \cite{fipa2004fipa}. Figure \ref{figure1} present the abstract bird-eye view of the envisioned self-organizing network, followed by the details and specification of each node in the forth-coming subsections.

\subsection{Incentive management mechanism based on escrow and smart contract} 
\label{sec:intution}
\subsubsection{Escrow}
Escrow is generally considered as an agreement of trust among two stakeholders formalizing a business service deal among one another by incorporating a trusted third party mediator in-between. Conventionally, escrows conducted by the administrative involvement of a trustee third-party which is responsible for making sure non of the stakeholders tangled in the business deal is defaulting. The resource n service consumers or generally buyer deposits a negotiated amount of money in the escrow and demands the service to be done in a given time. The resource n service provider or seller is notified of the deal and the along with a schedule time period complying the service needs to be performed in that amount of time. Thus, the seller is duty-bound for completing the service within the negotiated time for the negotiated charge at an acceptable quality. During the entire service time period, it is the guaranteed responsibility of the trusted third-party to ensure the security of the money. Once the seller completely performs the service and lets the trusted third-party escrow know the status of the deal. Next, it is the responsibility of the service consumer to approve/disapprove the service performed by the seller before the stipulated contract period ends. If the service performed is approved by the service consumer, the trusted third-party escrow will pay out the deposited money to the seller keeping a pre-decided percentage of the money as an escrow fee charge. If the service is done is disapproved by the service consumer, the escrow will hold the money unless a decision is taken by either of the stakeholders (the seller improves the quality of service / both the stakeholders decides not to go forward with the deal). Otherwise, at a time when an ongoing escrow has not ended, there might be a case that the seller does not want to go further with the contract; then the seller requests a cancellation of the escrow. This awaits cancellation confirmation from the service consumer upon whose positive response the trusted third-party escrow contract is canceled and the money deposited in the escrow is returned back to the service consumer keeping the pre-decided percentage of the money as the escrow fee charge. If the seller is unable to complete the service in the given time period, the deposited money is transferred back to the service consumer with the escrow fee being charged.

\begin{quote}
    Keeping the escrows model in view and framing it upon the incentive problem in self-organizing networks where it is possible for a resource consumer to extends its total operational capacity by subcontracting additional resources from resource providers making the federation incentive-driven.  
\end{quote}

\subsubsection{Trust Issues of Escrows with the proposed model}
The entrusted third-party intended to be the custodian for the escrow might not be honest - result in scenarios of unwanted troubles, contract manipulations,  theft, and collusion.

If any of the stakeholders is not truthful in the deal - and might result in locking up of one’s money in the escrow until the issue is resolved by involvement of legal authority.

Now here in our formalized self-organizing network, we ignore the case of physical escrow, as the scale of networks grows the physical escrows seem to be impractical for a two-node network.

\subsubsection{Blokchain Powered Decentralised Escrows}
Decentralized escrow refers to an escrow whose operations are not controlled by a trusted third-party, rather being transparent enough to be visible to everyone or intended to visible to everyone in the blockchain network. Ethereum \cite{wood2014ethereum} is such a decentralized blockchain network that allows transactions to be carried out between any two stakeholders without the need for a centralized third-party intermediary \cite{wood2014ethereum}. In addition, unlike Nakamoto's Bitcoin blockchain network \cite{nakamoto2008bitcoin}, Ethereum network provision the execution of code powered by gas. Hence, originating the concept of \textbf{smart contracts} that is a job-specific code written to ensure a particular job is performed in a decentralized constellation without the involvement of an intermediate trusted third-party.

Consequently, escrows can be modeled into fitting use-case of a blockchain-powered decentralized network to perform transactions. The central idea is to utilize smart contract for taking care of the security deposit and prevent the business stakeholders to default. Escrow over ethereum makes sure the service consumer’s money is not fiddled with, the seller gets the resource/service charge he demands, and the service consumer gets the resource/service he demands. This guarantee will ensure an incentive for the resource/service providers to be part of the self-organizing network and provide services to the network while compensating for the services being offered to the network.

\subsubsection{Escrow over Ethereum Smart Contract}
The escrow smart contract is presented as Appendix \ref{sec:appendixA} . The escrow smart contract is owned by an escrow owner (authority node) who is the one who creates the contract, formalized in the smart contract constructor. There’s a variable EscrowStatus that provides and tracks the current status of the escrow which will be set to unInitialized by default.

The escrow owner will initialize the contract for the two stakeholders i.e. a resource consumer and a resource provider by calling the Initialize() function of the contract while passing consumer address, provider address, fee percentage, and a final block number denoting the service time deadline. The EscrowStatus is set to initialized after the initialization step. It’s ensured that none of the addresses of the consumer or provider is equal to that of the escrow owner/authority node.

Once the escrow is initialized, the resource consumer can make any number of deposits to the contract using DepositInEscrowByConsumer() contract function calls which emit events for the same. Note that, the money deposited is owned by the contract and the escrow owner has no control over it, solving the problem of trust aroused due to dishonest third-party. After the deposit, the EscrowStatus will now hold the status of consumerDeposited.

The service time passes are measured as transactions take place in the network, and the latest block number increases a service time epoch passes. Once the service is provisioned by the resource provider before the latest block number exceeds the given block number limit, the resource provider approves the escrow and marks the resource provisioning by him to complete.

The resource consumer next holds the resource up-til the leased time, and relinquish the resource to the resource provider after utilizing it. In case the service quality is acceptable, the escrow is approved by the resource consumer. Since both the consumer and provider have approved the escrow, EscrowStatus now changes to serviceApproved state. Next, the smart contract automatically initiates payment of fee charged to the escrow owner a value which is decided by the fee percentage. Next, the smart contract automatically initiates the payment of the remaining balance amount to the resource provider address. At this point, the EscrowStatus is now changed to escrowComplete.

In case the service quality is not acceptable, the resource consumer does not approve of the service. After further negotiation with the resource provider, the resource provider can re-provision or decide to cancel the escrow. If the consumer too cancels the escrow, the entire amount of money deposited into the escrow will be refunded back to the consumer with the escrow owner keeping a pre-decided amount of sum as escrow fee charge. At this point, the EscrowStatus is changed to escowCancelled. After this, the escrow is now in steady-state and is ready to conduct another escrow.

Lastly, the escrow can be ended only by escrow owner so that the contract is destructed. 

\subsection{Authority Node}
\begin{figure}[H]
    \centering
      \includegraphics[scale=0.7]{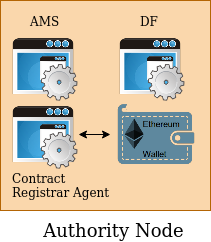}
  \caption{Component diagram of authority node}
  \label{figure2}
\end{figure}

We start our discussion on the working and internal of the proposed framework from the authority node because of the fact it is central to the working and understanding of the whole concept and mechanism of the proposed framework. The component diagram of the authority node is presented in Figure \ref{figure2}. 

The FIPA specifications state that a distributed multi-agent should have an Agent Management System (AMS) and a Directory Facilitator (DF) agent. The AMS agent represents the authority in the platform and has been tasked to control the distributed multi-agent platform and responsible for registering and destroying agents in the distributed agent platform and stopping the platform. Furthermore, the DF agent provides a directory that announces which agents are available on the platform, simply DF is a node lookup service. Due to these functionalities, it makes sense in our proposed framework the AMS and DF agents are kept on the authority node. Furthermore, multiple authority nodes can be made available to mitigate for the single-point of failure for a single authority node.

The contract registrar agent in the authority node is responsible for creating the escrow smart contract over the ethereum network, and the core of the business model of incentives anticipated through the proposed framework.  The communication between the contract registrar agent and the ethereum network will be facilitated through a Wallet console connected to the ethereum network.  The escrow smart contract between the resource consumer and the resource provider will be initialized by the contract registrar and on completion of a service transaction, the crypto-currency amount escrowed by the resource consumer node will be transferred into the resource provider wallet, along with a contract fee deducted and debited into the authority node account. The contract will be owned by the authority node to effectively imitate the physical escrow model and facilitate in arbitration and transaction rollbacks. The cyclic behavior according to the FIPA specification of the contract registrar agent is provided as Algorithm \ref{algorithm1}. The working of this behavior should be studied in conjunction with section \ref{subsubsection:contractBehavior}. Furthermore, the ethereum Smart Contract is provided as a supplementary attachment to this manuscript.


\begin{algorithm}[H]
 \caption{Contract Server}
 \small
 \begin{algorithmic}[1]
 \label{algorithm1}
 \renewcommand{\algorithmicrequire}{\textbf{Input: }}
 \renewcommand{\algorithmicensure}{\textbf{Output:}}
 \REQUIRE Wallet, ContractFee 
 \ENSURE  (BlockNo,TransactionHash)
  \STATE Message msg = ReceieveMessage()
  \STATE (ProviderAddress, ConsumerAddress, Deadline) =msg.getContent()
  \STATE Web3 = Connect to the Ethereum Network.
  \STATE Contract = Create Contract Instance;
  \STATE ContractAddress = Web3.deployContract(Contract,Wallet)
  \IF{$ContractAddress \neq \emptyset$}
  \STATE params = \{ProviderAddress, ConsumerAddress, ContractFee, Deadline\}
  \STATE Web3.callContractFunction(ContractAddress, "initEscrow", params, Wallet)
  \STATE  (BlockNo,TransactionHash) = WaitForTransactionToBeMined()
  \STATE Message response = CreateMessage(msg.getSender, FIPA.CONFIRM)
  \STATE response.setContent(ContractAddress)
  \STATE SendMessage(response)
   \ELSE
  \STATE Message response = CreateMessage(msg.getSender, FIPA.CANCEL)
  \STATE response.setContent(ContractAddress)
  \STATE SendMessage(response)
   \ENDIF
 
  \RETURN $ (BlockNo,TransactionHash)$
 \end{algorithmic}
 \end{algorithm}
 
The contract server behavior presented in Algorithm \ref{algorithm1} will execute cyclically until the lifespan of the contract registrar agent. The algorithm takes two inputs one is the ethereum wallet credentials, and the second one is the contract fee. The wallet credentials are used for interaction with the ethereum blockchain network. Line 1 of the algorithm opts to receive a FIPA request performative message from a resource consumer through the agent platform for contract initiation. The FIPA performative for agent communication is standardized as FIPA agent communication language message structure specification \cite{fipa2002fipa}. Line 2 extracts the message payload containing 3-tuple i.e., two ethereum network account addresses, and a contract initiation deadline. One of the received address is the address of a resource provider, while the second received address is the ethereum address of the resource consumer from which the message at Line 1 is received. Line 3 will allow the contract server to connect to the ethereum network required for deploying of the escrow smart contract and further interaction with the smart contract. Line 4 creates an instance of the escrow smart contract to push the contract into the ethereum network. Line 5 deploys the escrow smart contract on the ethereum network and will return the address of the contract. Line 5 is a blocking call and will return once the contract deployment transaction is mined in a block in the ethereum blockchain. Furthermore, the contract deployed in Line 5 will be in the ownership of the authority node to support the purposed business model through contract fee from the final transaction, and also to facilitates arbitration between the resource consumer or the resource provider due to their natural rationality.  Line 6 will check whether there is a contract address available to be sent to the requesting resource consumer. Line 7, initializes a parameter array that needs to be passed to the escrow smart contract initiation function. Line 8 uses the ethereum network interface to call the escrow initialization function of the resident escrow smart contract available at the address received on Line 5. Line 9, waits for the contract function call to be mined in the ethereum blockchain. Line 10-12 creates and sends a FIPA's confirm performative message to the resource consumer who requested the contract. If Line 6, yields false, meaning there is no contract address available i.e. the deployment of the contract is failed, then Line 14-16 creates and sends a FIPA's cancel performative message to inform the requesting resource consumer that the request can not be handled.

\subsection{Resource Consumers}
\label{sec:res_consumer}
\begin{figure}[H]
    \centering
      \includegraphics[scale=0.7]{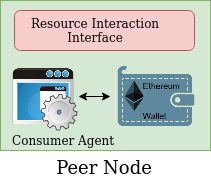}
  \caption{Component diagram of Resource Consumer}
  \label{figure3}
\end{figure}

The resource consumers are the lifeline behind the proposed business model. Resource consumers feed the resource providers and the authority nodes with the ethereum crypto-currency upon consuming a service/resource. The resource consumers in the case of a SOC will be a peer desktop computer node connected to internet consuming resources/services from other peer desktop computing node in the SOC network. In the case of federated cloud, a resource consumer is a data center/cloud service provider wishing to extend its total operational capacity by subcontracting additional resources from collaborating resources providing data centers. Similarly, in the case of ad-hoc/D2D networks, a resource consumer is a device wishing to extend its computer or communication capacity by using other resource provider devices in the network. The component diagram of a resource consumer is presented in Figure \ref{figure3}. The resource consumer is equipped with a FIPA compliant consumer agent, a resource interaction interface, and an ethereum wallet connected to the ethereum network. The consumer agent interacts with the ethereum network and ethereum escrow smart contract using the wallet connected to the ethereum network. The resource interaction interface is the northbound interface to the end-user or other software agents/systems if any, used for interacting with the resource acquired from the self-organizing network.

The consumer agent in the resource consumer nodes implement the following FIPA complaint agent behaviors to affirm the proper incentive management protocol in the proposed framework:

\subsubsection{Request Resource Behavior}`
\label{rrb}
Request resource behavior performs the service/resource request function and interaction with all the resource provider to find the best match resource/service in the self-organizing network. The request resource behavior is presented as Algorithm \ref{algorithm2}.

\begin{algorithm}[H]
 \caption{Request Resource Behavior}
 \small
 \begin{algorithmic}[1]
 \label{algorithm2}
 \renewcommand{\algorithmicrequire}{\textbf{Input: }}
 \renewcommand{\algorithmicensure}{\textbf{Output:}}
 \REQUIRE TargetResource 
 \ENSURE  (BestResource, BestProvider)
  \STATE array ProviderList = DFService.findAll("resource-provider")
  \STATE Message msg = CreateMessage(FIPA.CFP)
  \STATE msg.setContent(TargetResource)
  \STATE SendMessage(ProviderList,msg)
  \STATE NoOfReplies=0, BestScore=0
  \STATE BestProvider=$\emptyset$, BestResource=$\emptyset$
  \WHILE {$NoOfReplies < ProviderList.length$}
  \STATE Message response = ReceieveMessage("PROPOSE")
  \STATE (MatchScore,OfferredResource) = response.getContent()
  \IF{$MatchScore > BestScore$}
    \STATE BestScore = MatchScore
    \STATE BestProvider = response.getSender()
    \STATE BestResource = OfferredResource
  \ENDIF
  \STATE NoOfReplies++
  \ENDWHILE
  \RETURN $ (BestResource, BestProvider) $
 \end{algorithmic}
 \end{algorithm}
 
This behavior presented as Algorithm \ref{algorithm2} will be executed once an end-user i.e. a resource consumer triggers its execution and provides a target resource required. The required resource is abstracted as a resource vector in the form of $TargetResource:<s_1,s_2,...s_i>$, where $s_1...s_i$, is the specification of the required resource. According to the specified target resource the Algorithm \ref{algorithm2} starts with Line 1 by requesting the directory facilitation service to provide the list of resource providers who have registered a "resource-provider" service description with the agent platform. Line 2-4 of Algorithm \ref{algorithm2}, creates and sends a FIPA's call for proposal performative message to resource providers. The message is encapsulated with the required resource vector. Line 5-6 initializes some status variables to track the "propose" FIPA performative message responses from the resource providers to find the best match for the resource requested by the consumer. The while loop at Line 7, will remain true until and unless the response from all the resource providers is not received. Line 8-15, provides the implementation to select the best match resource from all of the resource providers. The match score is the similarity score between the requested resources and the resource available with the resource provider (this functionality is explained in section \ref{rhb} in detail). A resource provider will send his best-matched resource to the consumer, which is received here at the consumer on Line 8 of this algorithm. The if statement and its body starting from Line 10, allow this consumer to select the best resource available from all of the resource providers. Finally, the behavior returns the best resource hosted by a resource provider in the network. This returned output will be fed into forth mentioned consumer agent behaviors.

\subsubsection{Contract Behavior} 
\label{subsubsection:contractBehavior}
Contract Behavior is responsible to communicate with the contract registrar agent residing in the authority node to start an escrow smart contract with the resource provider.  This resource consumer's contract behavior is presented as Algorithm \ref{algorithm3}. The working of this behavior can be better understood in conjunction with contract server behavior presented as Algorithm \ref{algorithm1}.


\begin{algorithm}[H]
 \caption{Contract Behavior}
 \small
 \begin{algorithmic}[1]
 \label{algorithm3}
 \renewcommand{\algorithmicrequire}{\textbf{Input: }}
 \renewcommand{\algorithmicensure}{\textbf{Output:}}
 \REQUIRE BestResource, BestProviderAddress, Deadline,LeaseTime, Wallet 
 \ENSURE  (BlockNo,TransactionHash)
  \STATE Message msg = CreateMessage(FIPA.REQUEST)
  \STATE msg.setContent(BestProviderAddress,MyAddress,Deadline)
  \STATE SendMessage(AuthorityNode.ContractRegistrarAgent,msg)
  \STATE Message response = ReceieveMessage()
  \IF{response.checkPerformative("CONFIRM") $response.getContent() \neq \emptyset$}
  \STATE ContractAddress = response.getContent()
  \STATE Web3 = Connect to the Ethereum Network.
  \STATE depositAmount = LeaseTime $\times$ BestResource.price
  \STATE params = \{depositAmount\}
  \STATE Web3.callContractFunction(ContractAddress, "depositToEscrow", params, Wallet)
  \STATE  (BlockNo,TransactionHash) = WaitForTransactionToBeMined()
  \IF{$BlockNo,TransactionHash \neq \emptyset $}
  \RETURN $(BlockNo,TransactionHash, ContractAddress)$
  \ENDIF
  \RETURN $\emptyset$
  \ENDIF
 \end{algorithmic}
 \end{algorithm}

This contract behavior presented as Algorithm \ref{algorithm3} will be executed by the resource consumer agent automatically once it received the best resource and the resource provider address from the request resource behavior i.e. Algorithm \ref{algorithm2}. Consequently, lines 1-3 of this contract behavior send these details of a resource provider address, the address of this resource consumer, and a contract initiating deadline to the contract registrar agent residing on the authority node. The deadline is programmed in the smart contract to ensure the contract can only be made and initialized in a qualified time-interval specified by the resource consumer. These escrow smart contract initialization parameters sent by this contract behavior is received on line 2 and utilized at line 8 of the Algorithm \ref{algorithm1}. Furthermore, lines 10-11 at  Algorithm \ref{algorithm1} of the contract server behavior sends the ethereum public address of the smart contract which was here at this behavior received at line 4. The if statement on line 5 tests whether a contract address is received by checking the confirm message performative, if no contract address is being sent from the Algorithm \ref{algorithm1} this behavior will return null, specified as the else clause at line 13-14. Line 6-10 of this contract behavior handles the received contract address, to deposit ethereum crypto-currency into the escrow smart contract residing over the received contract address in the ethereum network. The deposit amount depends upon the valuation of the best resource, along-with the resource lease time that is provided as input to this behavior. The lease time and deadline are provided by the end-user, or it can be fixed depending upon the application. Once the transaction at line no. 10 is mined, line 11 will receive the block no and the transaction hash of the mined block and return it to the consumer agent for further usage in the next coming behavior.

\subsubsection{Acquire Resource Behavior} 
\label{subsubsection:AcquireResourceBehavior}
Acquire resource behavior is responsible to communicate with the resource provider from which the best resource was selected in Algorithm \ref{algorithm2}.  This acquire resource behavior is presented as Algorithm \ref{algorithm4}.

\begin{algorithm}[H]
 \caption{Acquire Resource Behavior}
 \small
 \begin{algorithmic}[1]
 \label{algorithm4}
 \renewcommand{\algorithmicrequire}{\textbf{Input: }}
 \renewcommand{\algorithmicensure}{\textbf{Output:}}
 \REQUIRE BestResource, BestProvider, ContractAddress, LeaseTime 
 \ENSURE  Resource Interface Details
  \STATE Message msg = CreateMessage(FIPA.ACCEPT\_PROPOSAL)
  \STATE msg.setContent(BestResource, ContractAddress, LeaseTime)
  \STATE SendMessage(BestProvider,msg)
  \STATE Message response = ReceieveMessage()
  \IF{response.checkPerformative("INFORM")}
  \RETURN $(Resource Interface Details)$
  \ELSE
  \RETURN $\emptyset$
  \ENDIF
 \end{algorithmic}
 \end{algorithm}

Once the consumer agent receives a valid contract address by executing the contract behavior (Algorithm \ref{algorithm3}), this acquire resource behavior will be executed. This behavior takes two input variables i.e. $BestResource, BestProvider$ returned by Algorithm \ref{algorithm2}, one user-provided inputs $LeaseTime$, and a $ContractAddress$ returned by Algorithm \ref{algorithm3} . This acquire resource behavior informs the selected resource provider about the escrow deposit made into the escrow contract. Line 1-3, convey this information to the selected resource provider, by sending a FIPA's accept proposal performative message accepting the proposal sent by the resource provider and received at line 8 of Algorithm \ref{algorithm2}.  The resource provider will check the deposit amount based on the leased time and the stipulated price of the resource under the escrow smart contract. If the escrow amount is available in the contract account, the resource provider will provide the resource interface to this resource consumer using a FIPA inform performative. Line 5-6 handles this message to be provisioned to the end-user. 

Once the user receives the resource interface details the lease time timer will start after a time threshold. Note that this is a proof of concept paper, this lease time timer is created to imitate the time event on which the resource release behavior can be triggered.


\subsubsection{Release Resource Behavior} 
\label{subsubsection:ReleaseResourceBehavior}
Release resource behavior is responsible to release the resource and interact with the escrow smart contract to release the balance into the resource provider account. Once the lease timer expires Algorithm \ref{algorithm5} will be called in automatically.

\begin{algorithm}[H]
 \caption{Release Resource Behavior}
 \small
 \begin{algorithmic}[1]
 \label{algorithm5}
 \renewcommand{\algorithmicrequire}{\textbf{Input: }}
 \renewcommand{\algorithmicensure}{\textbf{Output:}}
 \REQUIRE Resource, Provider, ContractAddress, Wallet 
 \ENSURE  null
\STATE Web3 = Connect to the Ethereum Network.
\STATE Web3.callContractFunction(ContractAddress, "approveEscrow", null, Wallet)
  \STATE Message msg = CreateMessage(FIPA.DISCONFIRM)
    \STATE msg.setContent(Resource)
  \STATE SendMessage(Provider,msg)
 \end{algorithmic}
 \end{algorithm}
 
This release resource behavior takes four input variables i.e. $Resource, Provider, ContractAddress, Wallet$. $Resource$ represents the resource that needs to be released, $Provider$ is the provider from which the resource was leased, $ContractAddress$ is the address of the escrow smart contract, and $Wallet$ is the ethereum wallet credentials of the consumer. Line 1-2 of Algorithm \ref{algorithm5} approves the escrow so that funds should be released to the resource provider. Line 3-5, inform the resource provider about the disengagement.

\subsection{Resource Providers}
The resource providers are the driving force behind the proposed business model. Resource providers provides its available resources from the resource pool to the resource consumers in the self-organizing overlay. The resource providers in the case of a SOC will be a peer desktop computer node connected to internet providing virtual resources/services to other peer desktop computing nodes i.e. resource consumers in the SOC network. In the case of federated cloud, a resource provider is a data center/cloud service provider wishing to minimizes its total operational cost by providing its available un-utilized resources to collaborating resources consuming data centers. Similarly, in the case of ad-hoc/D2D networks, a resource provider is a device wishing to provides its computer or communication capacity to be utilized by resource consumers devices in the network. As earlier mentioned in section \ref{sec:rw} Edinger et. al \cite{edinger2019money} hypothesizes "monetary  incentives  can enhance  the  intrinsic  motivation  for resource providers to  share  resources when  sharing takes  place  amongst  anonymous  users". Following this, blockchain based crypto-currency \cite{nakamoto2019bitcoin}  is a choice having a market value for the resource providers to enhance their motivation.

\begin{figure}[H]
    \centering
      \includegraphics[scale=0.7]{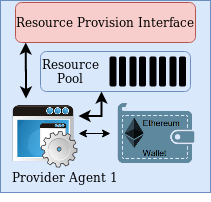}
  \caption{Component diagram of Resource Providers}
  \label{figure4}
\end{figure}

The component diagram of a resource provider is presented in Figure \ref{figure4}. The resource provider is equipped with a FIPA compliant provider agent, a resource provision interface, a resource pool from which resources are offered on per use basis,  and an ethereum wallet connected to the ethereum network. The provider agent interacts with the ethereum network and ethereum escrow smart contract using the wallet connected to the ethereum network. The resource provision interface is the northbound interface to the resource interaction interface of the resource consumers or any other software agents/systems if any, used by the resource consumers for interacting with the resource acquired from the self-organizing network.

The providers agent in the resource provider nodes implement the following FIPA complaint agent behaviors to affirm the proper incentive management protocol in the proposed framework:

\subsubsection{Request Handle Behavior} 
\label{rhb}
This request handle behavior is cyclic and is continuously running to accept resource request from the resource consumers. The behavior is initialized when the provider agent starts up. The request handle behavior is presented as Algorithm \ref{algorithm6}.  Algorithm \ref{algorithm6}  listens for the CFP messages from the resource consumers request resource behavior presented as Algorithm \ref{algorithm2} and respond it accordingly.

\begin{algorithm}[H]
 \caption{Request Handle Behavior}
 \small
 \begin{algorithmic}[1]
 \label{algorithm6}
 \renewcommand{\algorithmicrequire}{\textbf{Input: }}
 \renewcommand{\algorithmicensure}{\textbf{Output:}}
 \REQUIRE catalogue
 \ENSURE  null
\STATE 			Message msg = ReceiveMessage();
\IF{msg.checkPerformative("CFP")}
\STATE 	requestedResource = msg.getContentObject();
\STATE  bestMatch = null;
\STATE	bestSimilarityScore = 0.0
\STATE	currentSimilarityScore = 0.0
\STATE  ResourcesList = catalogue;
\FOR{\textbf{each} res \textbf{in} ResourcesList}
\STATE	currentSimilarityScore = CosineSimilarity(requestedResource, res)
\IF{$currentSimilarityScore  > bestSimilarityScore$} 
\STATE	bestSimilarityScore = currentSimilarityScore
\STATE	bestMatch = res
\ENDIF
\ENDFOR

\IF{bestMatch != null} 
\STATE	reply = msg.createReply();
\STATE					reply.setPerformative(PROPOSE);
\STATE					reply.setContentObject(bestMatch);
\ELSE		
\STATE	reply = msg.createReply();
\STATE					reply.setPerformative(REFUSE);
\STATE					reply.setContent("not-available");
\ENDIF				
\STATE  			SendMessage(reply);
  \ELSE
\STATE  			Do Nothing

  \ENDIF
 \end{algorithmic}
 \end{algorithm}

Algorithm \ref{algorithm6} takes as an input the resource catalog the resource provider is offering right now. The catalog is essentially a list of resources ( a resource is represented as a row vector as mentioned in section \ref{rrb}). Line 1 of Algorithm \ref{algorithm6} opts to receive a message and check its FIPA's CFP performative in Line 2. Line 3 extracts the requested resource vector sent by a resource consumer. Line 4-7, initialize some status variables to lookup for the best matching resource in the resource catalog against the requested resource received in Line 3.  The loop at line 8-14 calculates the cosine similarity score between the requested resource and a resource from the resource catalog in every iteration and records the best-matched resource and its score in Line 10-13. Line 9 uses cosine similarity for comparing the two resource vectors, and is defined as follows:

\begin{equation}
\cos ({\bf t},{\bf e})= {{\bf t} {\bf e} \over \|{\bf t}\| \|{\bf e}\|} = \frac{ \sum_{i=1}^{n}{{\bf t}_i{\bf e}_i} }{ \sqrt{\sum_{i=1}^{n}{({\bf t}_i)^2}} \sqrt{\sum_{i=1}^{n}{({\bf e}_i)^2}} }
\end{equation}

Where $t$ and $e$ are  resource vector in the form of$<s_1,s_2,...s_n>$, where $s_1...s_n$, is the specification of the resource. The use of similarity function makes things simple and realistic as every resource provider will have to use the same similarity function. Furthermore, one can argue why not to use the price as the matching criteria, price is just one factor in the resource vector there are other resource specification involved which should also need to be included in the best match decision and cosine similarity seems good for such type of row vector comparisons. Line 15-18 will check and prepare a message containing the best-matched resource specifications using a FIPA's propose performative to the resource consumer which requested the resource.  The else statement of Line 19-22 prepares a message using a FIPA's refuse performative to the resource consumer which requested the resource to specify the refusal of the request. Line 24 sends this response message for both the cases. The propose performative sent by the resource provider is handled at Algorithm \ref{algorithm2} line 8-15 of the resource consumer.

\subsubsection{Lease Resource Behavior} 
This lease resource behavior just like the previous behavior is cyclic and is continuously running for responding to the proposals sent by the resource consumers in response to the proposals sent by the request handle behavior in Algorithm \ref{algorithm6}. The lease resource behavior is presented as Algorithm \ref{algorithm7}.

Algorithm \ref{algorithm7} takes input an ethereum wallet information and two data structures from the resource provider agent one is the resource catalog and the second one is the consumer resource map to record the resource lease. Line 1 receives a message from resource consumers. Line 2 checks for an accept proposals FIPA's performative message which is sent by Algorithm \ref{algorithm4} of the resource consumer. Line 3, extracts the information about the resource which needs to be leased to the requesting resource consumer. Line 4 gets this resource from the resource catalog. Line 5 will check whether the resource was successfully popped from the resource pool, otherwise due to any reasons the resource is unavailable in the catalog a FIPA failure performative message is prepared and sent to the resource consumer in Line 17-20. Line 6 puts the record of the lease in key-value map specifying which consumer has taken which resource. Line 7, gets the ethereum contract address of the smart contract in which the escrow amount is being deposited by the resource consumer. Line 8-11 connects with the ethereum wallet and approves the escrow enabling the contract. Line 12 -15 will send a FIPA's inform performative message enclosing the resource interaction details addressed toward the requesting resource consumer.

\begin{algorithm}[H]
 \caption{Lease Resource Behavior}
 \small
 \begin{algorithmic}[1]
 \label{algorithm7}
 \renewcommand{\algorithmicrequire}{\textbf{Input: }}
 \renewcommand{\algorithmicensure}{\textbf{Output:}}
 \REQUIRE catalogue, ConsumerResourceMap, Wallet
 \ENSURE  null
\STATE 			Message msg = ReceiveMessage();
\IF{msg.checkPerformative("ACCEPT\_PROPOSAL")}
  
\STATE requestedResource = msg.getContentObject();
\STATE				poppedResource = catalogue.remove(requestedResource);
\IF{poppedResource != null}
\STATE  	ConsumerResourceMap.put(msg.getSender(), poppedResource); 
\STATE  	contractAddress = requestedResource.getContractAddress();

 \STATE Web3 = Connect to the Ethereum Network.
  \STATE params = \{\}
  \STATE Web3.callContractFunction(contractAddress, "approveEscrow", params, Wallet)
  \STATE  (BlockNo,TransactionHash) = WaitForTransactionToBeMined()

\STATE  				Message reply = msg.createReply();
\STATE  				reply.setPerformative(INFORM);
\STATE  			reply.setContent("Resource Interaction Details");
\STATE  				SendMessage(reply);

  \ELSE
\STATE  					Message reply = msg.createReply();
\STATE  					reply.setPerformative(FAILURE);
\STATE  					reply.setContent("not-available");
\STATE  				SendMessage(reply);

 \ENDIF
  \ELSE
\STATE  			Do Nothing

  \ENDIF
 \end{algorithmic}
 \end{algorithm}

\subsubsection{Release Resource Behavior} 
This release resource behavior is the resource provider version of the releasing resource which works according to the perspective of the resource provider and will interact with the resource consumer resource release behavior presented in section \ref{subsubsection:ReleaseResourceBehavior}. The resource provider release resource behavior is presented as Algorithm \ref{algorithm8}.

Algorithm \ref{algorithm8} takes the catalog and consumer resource map as parameters. Line 1 reads a message from a resource consumer through the underlying message platform. Line 2 checks for a FIPA's disconfirm performative. If not successful will prepare and send a FIPA failure performative message in Line 11-14. If line 2 yields true then, line 3 will extract the information of the resource that a resource consumer wants to release. Line 4, will verify there is a resource that needs to be released. Line 5, create a reply message template for the requesting resource consumer. Line 6 puts the resource back in the catalog, making it available to other resource consumers to lease it. Line 7, remove the lease record maintained by the resource provider agent. Line 8 set the reply message performative to FIPA's disconfirm to tell the requesting resource consumer agent about the dis-engagement. Line 9 finally sends the message to the requesting resource consumer.

\begin{algorithm}[H]
 \caption{Release Resource Behavior}
 \small
 \begin{algorithmic}[1]
 \label{algorithm8}
 \renewcommand{\algorithmicrequire}{\textbf{Input: }}
 \renewcommand{\algorithmicensure}{\textbf{Output:}}
 \REQUIRE catalogue, ConsumerResourceMap
 \ENSURE  null
 
 \STATE 			Message msg = ReceiveMessage();
\IF{msg.checkPerformative("DISCONFIRM")}

 \STATE releasedResource = msg.getContentObject();
\IF{releasedResource != null}
\STATE  				Message reply = msg.createReply();
				
\STATE  				catalogue.put(releasedResource.getResourceName(), releasedResource);
\STATE  					ConsumerResourceMap.remove(msg.getSender(), releasedResource); 
\STATE  					reply.setPerformative(DISCONFIRM);
\STATE  				SendMessage(reply);  
 \ELSE
 \STATE  				Message reply = msg.createReply();
\STATE  					reply.setPerformative(FAILURE);
\STATE  					reply.setContent("not-available");
\STATE  				SendMessage(reply);  

\ENDIF

  \ELSE
\STATE  			Do Nothing

  \ENDIF
 
 \end{algorithmic}
 \end{algorithm}
 
\section{Proof of Concept}
\label{sec:poc}
\begin{figure*}
    \centering
      \includegraphics[scale=0.5]{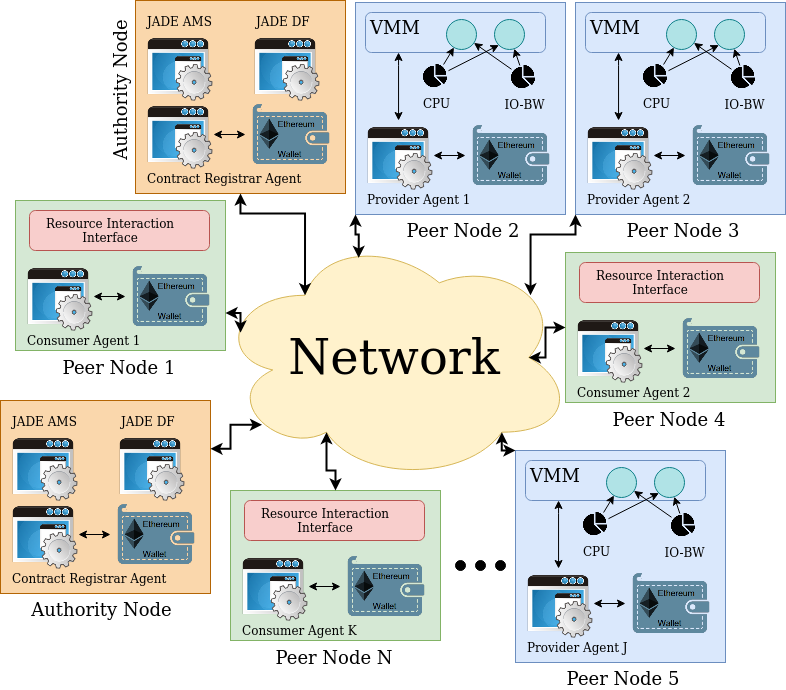}
  \caption{Topology and Component Details of Proof of Concept Implementation}
  \label{figure5}
\end{figure*}

For the proof of concept of the proposed framework, we developed a simulated testbed and utilize off the shelf technologies to conceptualize the Ethereum blockchain-powered incentive management solution for a self-organizing cloud. Figure \ref{figure5} presents the topology of the proposed implementation along with the component details of each of the nodes. This proof of concept emulates the self-organizing cloud environment where desktop computers integrating volunteer computing and peer-to-peer (P2P) networking into cloud architectures anticipating an architecture of a gigantic self-organizing cloud (SOC) to reap the huge potential of untapped commodity computing power over the Internet \cite{6200263}. Figure \ref{figure5} clearly presents the resource providers agents interacting with the on-device VMM to trade and provider virtual resources to the resource consumers in the self-organizing network.

The proposed framework was formalized using agent-based methodologies, ethereum blockchain smart contracts, and cryptocurrency. The agent methodologies is incorporated by using the java agent development framework (JADE) \cite{bellifemine2005jade} \footnote{\url{https://jade.tilab.com/}}. JADE serves as a building block for the proposed framework implementation.  JADE is fully implemented in the Java language which also implies that our proposed framework proof of concept implementation is also fully in Java language. JADE simplifies the implementation of multi-agent systems through a middle-ware that complies with the FIPA specifications and through a set of tools that support the debugging and deployment phases. A JADE-based system can be distributed across machines (which does not even need to share the same OS) and the configuration can be controlled making JADE perfect for implementing a self-organizing system. The configuration can be even changed at run-time by moving agents from one machine to another, as and when required. Lastly, we have configured our own private ethereum test net in order to verify the proposed framework.

\subsection{Experimental Setup}
The proof of concept implementation is performed on an Intel(R) Core(TM) i7-7700 CPU @ 3.60GHz having 16 GB RAM configured with Ubuntu 18.04 LTS OS. The ethereum test-net was created using Go Ethereum blockchain implementation. The proposed framework primarily focused on solving the incentive problems in order to motivate nodes to join the self-organizing network and provide their resources to the network. The motivation was increased by monetizing the resource offered by the resource providers using the ethereum smart contract and cryptocurrency. In order to achieve this, the resource providers and consumers are simulated through software agents implemented using JADE to provide a resource and consume a resource. These software agents can readily be plugged in a physical testbed all because of the JADE middle-ware.

\subsection{Resource Providers Agents View}
The resource provider agent when it starts-up loads five random resources from a resource trace file presented as Table \ref{tab1}. These five resources will be added to the resource pool of the resource providers and once a request from a resource consumer reaches the resource provider the request is matched with these resources available in the resource pool. 

\begin{table}[]
\caption{Trace File for resource providers}
\label{tab1}
\begin{tabular}{@{}lllllll@{}}
\toprule
Title & Wei\slash hr & MIPS\footnote{Million of Instructions Per Second} & \$ \slash GB & RAM & BW (MBPS) & CPU Cores \\ \midrule
t3a.nano     & 0.0047        & 4744   & 0.284       & 0.5      & 18            & 2           \\
t3.nano      & 0.0052        & 7500   & 0.276       & 0.5      & 82            & 2           \\
t2.nano      & 0.0058        & 8800   & 0.32        & 0.5      & 44            & 1           \\
t3a.micro    & 0.0094        & 9900   & 0.352       & 1        & 4             & 2           \\
t3.micro     & 0.0104        & 10000  & 0.096       & 1        & 66            & 2           \\
t2.micro     & 0.0116        & 10240  & 0.116       & 1        & 31            & 1           \\
t3a.small    & 0.0188        & 13800  & 0.044       & 2        & 59            & 2           \\
t3.small     & 0.0208        & 14000  & 0.092       & 2        & 55            & 2           \\
t2.small     & 0.023         & 18938  & 0.036       & 2        & 12            & 1           \\
a1.medium    & 0.0255        & 19200  & 0.26        & 2        & 40            & 1           \\
t3a.medium   & 0.0376        & 27079  & 0.12        & 4        & 35            & 2           \\
t3.medium    & 0.0416        & 42820  & 0.268       & 4        & 68            & 2           \\
t2.medium    & 0.0464        & 49161  & 0.064       & 4        & 23            & 2           \\
a1.large     & 0.051         & 49360  & 0.284       & 4        & 100           & 2           \\
t3a.large    & 0.0752        & 53840  & 0.04        & 8        & 23            & 2           \\
t3.large     & 0.0832        & 59455  & 0.28        & 8        & 24            & 2           \\
m5a.large    & 0.086         & 65770  & 0.008       & 8        & 57            & 2           \\
t2.large     & 0.0928        & 71120  & 0.172       & 8        & 12            & 2           \\
m5.large     & 0.096         & 78440  & 0.04        & 8        & 96            & 2           \\
m4.large     & 0.1           & 82300  & 0.4         & 8        & 79            & 2           \\
a1.xlarge    & 0.102         & 83000  & 0.12        & 8        & 54            & 4           \\
m5ad.large   & 0.103         & 90749  & 0.048       & 8        & 93            & 2           \\
m5d.large    & 0.113         & 92100  & 0.256       & 8        & 44            & 2           \\
m5n.large    & 0.119         & 97125  & 0.332       & 8        & 99            & 2           \\
m5dn.large   & 0.136         & 106924 & 0.204       & 8        & 99            & 2           \\
t3a.xlarge   & 0.1504        & 113093 & 0.216       & 16       & 69            & 4           \\
t3.xlarge    & 0.1664        & 115625 & 0.036       & 16       & 95            & 4           \\
m5a.xlarge   & 0.172         & 117160 & 0.236       & 16       & 10            & 4           \\
t2.xlarge    & 0.1856        & 133740 & 0.368       & 16       & 9             & 4           \\
m5.xlarge    & 0.192         & 147600 & 0.164       & 16       & 72            & 4           \\
m4.xlarge    & 0.2           & 176170 & 0.06        & 16       & 10            & 4           \\
a1.2xlarge   & 0.204         & 238310 & 0.4         & 16       & 68            & 8           \\
m5ad.xlarge  & 0.206         & 304510 & 0.324       & 16       & 88            & 4           \\
m5d.xlarge   & 0.226         & 317900 & 0.38        & 16       & 93            & 4           \\ \bottomrule
\end{tabular}
\end{table}

\begin{figure}
    \centering
      \includegraphics[scale=0.33]{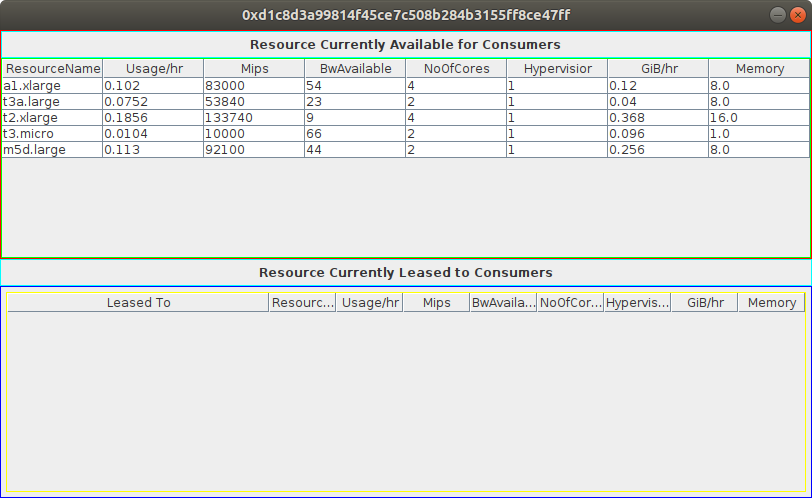}
  \caption{Resource Provider Agent GUI View}
  \label{figure6}
\end{figure}

The resource provider agent once loaded look like Figure \ref{figure6}. The upper grid in Figure \ref{figure6} represents the resources available in the resource pool which can be offered to the consumers in the self-organizing network. The lower grid in the GUI of the resource provider agents presents the resources that are currently being leased to the consumers. Besides this visual look out of the resource provider agent, the resource provider agent also has an ethereum wallet that is loaded from an Ethereum blockchain crypto JSON wallet key-store file. This wallet credential in the key-store file will represent the resource provider in the ethereum network and is used by the software agent to interface with the ethereum network, and the escrow smart contract. All the transactions will be channeled into the account specified in the key-store file.

\subsection{Resource Consumer Agents View}

\begin{figure}
    \centering
      \includegraphics[scale=0.5,angle=90]{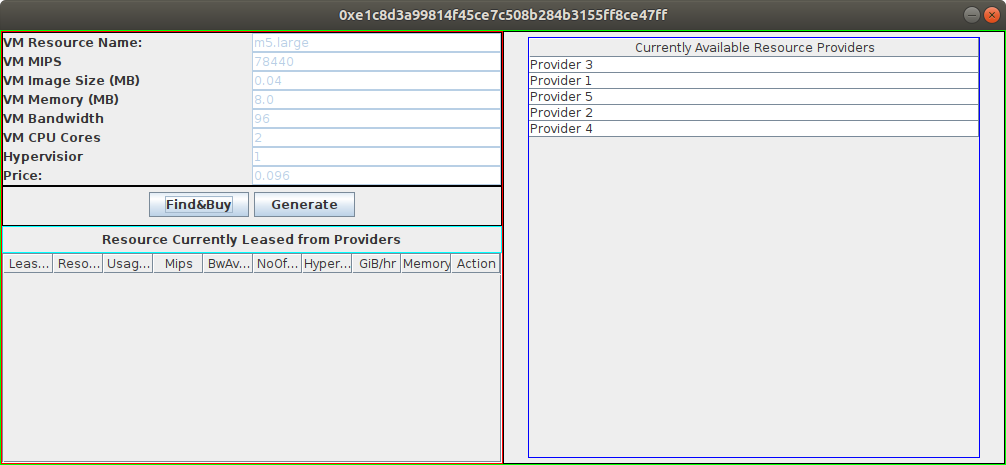}
  \caption{Resource Consumer Agent GUI View}
  \label{figure7}
\end{figure}

The resource consumer agent when starts-up loads a random resource from the resource trace file presented as Table \ref{tab1}. This resource would serve as the requesting resource from the self-organizing network. Figure \ref{figure7} represents the GUI of the resource consumer agent with configurable resource specifications, or you can just generate a random request from the trace file. Figure \ref{figure7} also presents the list of active providers from which the resource can be bought upon finding the best match using cosine distance as mentioned in section \ref{rhb}. The list of active providers is published by the directory facilitator agent residing on the authority node.

Similarly, like the resource provider agent, the resource consumer agent also has an ethereum wallet that is loaded from an Ethereum blockchain crypto JSON wallet key-store file to interface with the ethereum network, and the escrow smart contract. The per usage price of the resource will be deducted from the specified ethereum wallet. A Resource consumer can buy as many resources from the network as needed or a policy implementation can be made upon the type of self-organizing network the framework is being used.

\subsection{Contract Registrar Agent View}
\begin{figure}[H]
    \centering
      \includegraphics[scale=0.4]{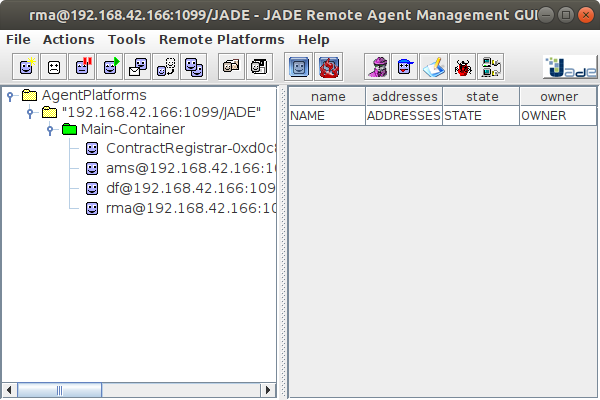}
  \caption{Authority Node GUI View}
  \label{figure8}
\end{figure}

The contract registrar agent resides on the authority node. Contract registrar agent the core of the business model also has an ethereum wallet which is loaded from an Ethereum blockchain crypto JSON wallet key-store file to interface with the ethereum network, and initialize the escrow smart contracts, upon consumer request, and then deduct the escrow fee from the transactions between the resource consumer and provider upon completion of the escrow smart contract. Figure \ref{figure8}, shows a JADE remote agent management GUI from the proof-of-concept implementation, showing a contract registrar agent being loaded along with directory facilitator and other JADE platform agents.

\subsection{Headless Simulation }

Besides the above proof of concept GUIs for each agent, we ran simulations in headless mode without GUIs to gather some statistics related to the transactions that occurred on the trade of resources. The proof of concept experiment executes five instances of resource providers agents imitating five resource providers. Every agent has his own set of to be offered resources in the resource pool, along with a different Ethereum blockchain crypto wallet effectively imitating the uniqueness of resource providers agent in the ethereum network. Similarly, every resource provider wallet will be debited once a resource consumer consumes a resource and the escrow smart contract terminates gracefully. Furthermore, we ran multiple settings of resource consumers agents in headless mode. For simplicity first consider the case of one resource consumer agent in the simulated network whose job is to imitate hold and consume of a resource for the next epoch in the simulation, while every other behavior runs as described in section \ref{sec:res_consumer}. An epoch is a hold period of the resource was fixed to one hour and after the imitating elapse of the epoch the resource consumer release the resource and complete the smart contract, this assumption simplify the matters for understanding/solving the motivation problem caused in self-organizing networks.

\begin{table*}[]
\caption{Simulation Trace: Single Resource consumer vs. five Resource providers}
\label{tab:Scenario1}

\centering
\large
\begin{tabular}{@{}llllllll@{}}
\toprule
Title & Wei\slash hr & MIPS\footnote{Million of Instructions Per Second} & \$ \slash GB & RAM & BW (MBPS) & CPU Cores & Cosine Score      \\ \midrule
\multicolumn{8}{l}{Resource Consumer Randomly requested}                         \\
t3a.small  & 0.0188 & 13800  & 0.044 & 2   & 59  & 2         &                   \\
\multicolumn{8}{l}{Resource Provider 1  Resource Pool}                           \\
t3a.micro  & 0.0094 & 9900   & 0.352 & 1   & 4   & 2         & 0.999992503425711 \\
m5.large   & 0.096  & 78440  & 0.04  & 8   & 96  & 2         & 0.999995336207502 \\
t3.nano    & 0.0052 & 7500   & 0.276 & 0.5 & 82  & 2         & 0.999977827484931 \\
m5a.large  & 0.086  & 65770  & 0.008 & 8   & 57  & 2         & 0.999994183623252 \\
t3a.small  & 0.0188 & 13800  & 0.044 & 2   & 59  & 2         & \textbf{1}                 \\
\multicolumn{8}{l}{Resource Provider 2 Resource Pool}                            \\
m5.xlarge  & 0.192  & 147600 & 0.164 & 16  & 72  & 4         & 0.999992819700923 \\
m5d.xlarge & 0.226  & 317900 & 0.38  & 16  & 93  & 4         & 0.999992055462973 \\
m5.large   & 0.096  & 78440  & 0.04  & 8   & 96  & 2         & \textbf{0.999995336207502} \\
m5a.large  & 0.086  & 65770  & 0.008 & 8   & 57  & 2         & 0.999994183623252 \\
m5dn.large & 0.136  & 106924 & 0.204 & 8   & 99  & 2         & 0.999994380189432 \\
\multicolumn{8}{l}{Resource Provider 3 Resource Pool}                            \\
m4.large   & 0.1    & 82300  & 0.4   & 8   & 79  & 2         & 0.999994495539404 \\
a1.medium  & 0.0255 & 19200  & 0.26  & 2   & 40  & 1         & 0.9999975923618   \\
t2.micro   & 0.0116 & 10240  & 0.116 & 1   & 31  & 1         & \textbf{0.999999218978273} \\
t3.nano    & 0.0052 & 7500   & 0.276 & 0.5 & 82  & 2         & 0.999977827484931 \\
m5d.xlarge & 0.226  & 317900 & 0.38  & 16  & 93  & 4         & 0.999992055462973 \\
\multicolumn{8}{l}{Resource Provider 4 Resource Pool}                            \\
t2.micro   & 0.0116 & 10240  & 0.116 & 1   & 31  & 1         & 0.999999218978273 \\
t3.small   & 0.0208 & 14000  & 0.092 & 2   & 55  & 2         & \textbf{0.999999939860052} \\
t3a.medium & 0.0376 & 27079  & 0.12  & 4   & 35  & 2         & 0.999995548858419 \\
a1.large   & 0.051  & 49360  & 0.284 & 4   & 100 & 2         & 0.999997462590219 \\
m5a.large  & 0.086  & 65770  & 0.008 & 8   & 57  & 2         & 0.999994183623252 \\
\multicolumn{8}{l}{Resource Provider 5 Resource Pool}                            \\
m5ad.large & 0.103  & 90749  & 0.048 & 8   & 93  & 2         & 0.99999470786012  \\
m5d.xlarge & 0.226  & 317900 & 0.38  & 16  & 93  & 4         & 0.999992055462973 \\
a1.2xlarge & 0.204  & 238310 & 0.4   & 16  & 68  & 8         & 0.999992030759267 \\
t3a.xlarge & 0.1504 & 113093 & 0.216 & 16  & 69  & 4         & 0.999993277084396 \\
t3.small   & 0.0208 & 14000  & 0.092 & 2   & 55  & 2         & \textbf{0.999999939860052} \\ \bottomrule
\end{tabular}
\end{table*}

Table \ref{tab:Scenario1} presents the scenario generated by the simulating a single epoch of a resource consumer vs. five resource providers. The resource consumer sends a call for proposal for his required resource to the resource providers in the network. Upon receiving the CFP request each of the providers will match the request with their resource pool using the cosine similarity function, and only send in response the best match back with requested a resource. The best match with each resource provider according to the requested posed the consumer is in bold-face at the cosine score column of Table \ref{tab:Scenario1}. Once these best match from the individual resource providers reaches the resource consumer selected the response of Resource Provider 1 and sends a request to the contract registrar agent to start an escrow contract and then put the designated amount of 0.0188 Wei into the smart escrow contract, and send an accept proposal message to the resource provider in question. Upon completion of the epoch, once the resource consumer releases the resource and finishes the smart contract, 2\% transaction fee is deducted by the contract registrar agent while the rest is being debited into the ethereum account of resource provider 1.

Now consider the case of simulating ten epochs of a resource consumer vs. five resource providers. Where in every epoch the resource consumer requested a different resource and after getting the best match resource from the network likewise, the resource consumer get into a smart escrow contract with the resource provider providing the best match resource, and upon completion of the epoch the resource is released and the amount held by the contract is transferred to the resource provider. Table \ref{tab:Scenario2} presents the simulation trace of ten epochs wherein each row the first column presents the simulation epoch. Whereas the requested resource column presents the resource the consumer requested from the network. The details specification of these resources is already defined earlier in Table \ref{tab1}. The winner column specifies the resource provider which has the best-matched resource for the request. The resource offered column specifies the resource which is being offered by the resource provider from its resource pool presented in Table \ref{tab:Scenario1}. The Wei column presents the ethereum cryptocurrency amount that was charged for the epoch which was held by the contract once the consumer starts a contract and then transferred the balance to the resource provider. The contract fee column specifies the benefit the contract registrar gets for every contract. In the ten epochs presented in Table \ref{tab:Scenario2} the total transactions in Wei's sum up to 1.1119, while the contract registrar agent received a total of 0.022238 Wei as 2\% contract fee in each epoch.

\begin{table}[]
\caption{Simulation Trace: Resource consumer request resource in ten epochs}
\label{tab:Scenario2}
\begin{tabular}{@{}llllll@{}}
\toprule
Epoch & \begin{tabular}[c]{@{}l@{}}Requested \\ Resource\end{tabular} & Winner                                                         & \begin{tabular}[c]{@{}l@{}}Resource \\ Offered\end{tabular} & Wei    & \begin{tabular}[c]{@{}l@{}}Contract \\ Fee (2\%)\end{tabular} \\ \midrule
1     & m5.xlarge                                                     & \begin{tabular}[c]{@{}l@{}}Resource \\ Provider 2\end{tabular} & m5.xlarge                                                   & 0.192  & 0.00384                                                       \\
2     & m5ad.large                                                    & \begin{tabular}[c]{@{}l@{}}Resource\\ Provider 5\end{tabular}  & m5ad.large                                                  & 0.103  & 0.00206                                                       \\
3     & m5d.large                                                     & \begin{tabular}[c]{@{}l@{}}Resource\\ Provider 2\end{tabular}  & m5.xlarge                                                   & 0.192  & 0.00384                                                       \\
4     & m5.large                                                      & \begin{tabular}[c]{@{}l@{}}Resource \\ Provider 1\end{tabular} & m5.large                                                    & 0.096  & 0.00192                                                       \\
5     & t3.micro                                                      & \begin{tabular}[c]{@{}l@{}}Resource\\ Provider 1\end{tabular}  & t3a.small                                                   & 0.0188 & 0.000376                                                      \\
6     & t3.medium                                                     & \begin{tabular}[c]{@{}l@{}}Resource\\ Provider 4\end{tabular}  & t3a.medium                                                  & 0.0376 & 0.000752                                                      \\
7     & t2.medium                                                     & \begin{tabular}[c]{@{}l@{}}Resource\\ Provider 2\end{tabular}  & m5.xlarge                                                   & 0.192  & 0.00384                                                       \\
8     & a1.medium                                                     & \begin{tabular}[c]{@{}l@{}}Resource\\ Provider 3\end{tabular}  & a1.medium                                                   & 0.0255 & 0.00051                                                       \\
9     & m5a.xlarge                                                    & \begin{tabular}[c]{@{}l@{}}Resource\\ Provider 5\end{tabular}  & a1.2xlarge                                                  & 0.204  & 0.00408                                                       \\
10    & a1.large                                                      & \begin{tabular}[c]{@{}l@{}}Resource\\ Provider 4\end{tabular}  & a1.large                                                    & 0.051  & 0.00102                                                       \\
      &                                                               &                                                                & Total Wei:                                                  & 1.1119 & 0.022238                                                      \\ \bottomrule
\end{tabular}
\end{table}

\begin{figure}[H]
    \centering
      \includegraphics[scale=0.8]{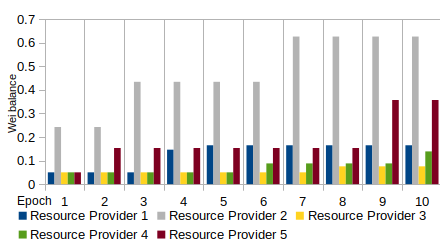}
  \caption{Wei Balance of Resource Providers}
  \label{figure9}
\end{figure}

Table \ref{tab:Scenario2} verify our claim as monetizing the resources provided by the resource providers, thus, improving their motivation to join the network, provide services, and get paid for it. This phenomenon is further explained in the graph presented as Figure \ref{figure9}.  Figure \ref{figure9} represents each epoch of Table \ref{tab:Scenario2} as x-axis, while the y-axis presents the Wei balance of a resource provider's Ethereum crypto wallet, assuming that in the start of the simulation every resource provider starts with 0.05 wei balance. Figure \ref{figure9} visualizes Table \ref{tab:Scenario2} presenting the increment in the wallet of resource providers at every epoch.


As a matter of understanding we have runs multiple resource consumers and multiple resource providers also in the simulated test-bed environment. However, for the sake of simplicity, brevity, and clarity we believe the motivation problem and its solution is better explained using single resource consumers and multiple epochs. Thus explaining how a resource provider gets benefited, as well as the contract registrar agent also gets benefits, thus opening a prominent blockchain-based business model.

\section{Conclusion}
\label{sec:conc}
This article proposes a blockchain-based incentive management framework for self-organizing networks. The proposed framework utilizes smart contract and agent-based methodologies to formalize a decentralized escrow between the resource consumer and resource provider. This novel way helps in increasing the motivation of nodes providing services to the network. Furthermore, the proposed framework opens a blockchain-based business model for entrepreneurs and outlet owners. Lastly, the proposed framework registers a whole new range of research problems to be studied ranging from a market place to apply auction/game theory or applying the framework on smart grids, or studying system's net utility maximization to name a few.

\appendices
\onecolumn
\section{Escrow Smart Contract}
\label{sec:appendixA}
This contract is conceptualized and recreated based from on open source Smart Escrow Implementation available at \footnote{\url{https://github.com/rounakdatta/escrow-dapp/blob/nodejsapp/contracts/escrow.sol}}. This smart contract stands true for solidity version 0.6.2 and requires the ethereum virtual machine support for this version of the contract specification. The source code of the escrow smart contract according to the business logic of our proposed model is presented as: 
\small
\begin{lstlisting}[numbers=left,basicstyle=\small]
pragma solidity ^0.6.2;
import "./SafeMath.sol";
contract Escrow {
    mapping (address => uint256) private EscrowAccountLedger;
    address payable public provider;
    address payable public consumer;
    address payable public authorityNode;
    uint256 public blockNumber;
    uint public feePercent;
    uint256 public escrowCharge;
    bool public providerApproval;
    bool public consumerApproval;
    bool public providerCancel;
    bool public consumerCancel;
    uint256[] public deposits;
    uint256 public feeAmount;
    uint256 public providerAmount;
    enum EscrowState { unInitialized, initialized, consumerDeposited,
    serviceApproved, escrowComplete, escrowCancelled }
    EscrowState public EscrowStatus = EscrowState.unInitialized;
    event Deposit(address depositor, uint256 deposited);
    event ServicePayment(uint256 blockNo, uint256 contractBalance);
    modifier onlyConsumer() {
        if (msg.sender == consumer) {
            _;
        } else {
            revert();
        }
    }
    modifier onlyAuthorityNode() {
        if (msg.sender == authorityNode) {
            _;
        } else {
            revert();
        }
    }    
    modifier checkBlockNumber() {
        if (blockNumber > block.number) {
            _;
        } else {
            revert();
        }
    }
    modifier ifApprovedOrCancelled() {
        if ((EscrowStatus == EscrowState.serviceApproved) ||
        (EscrowStatus == EscrowState.escrowCancelled)) {
            _;
        } else {
            revert();
        }
    }
    constructor () public {
        authorityNode = msg.sender;
        escrowCharge = 0;
    }
    fallback () external { // solhint-disable-line
        // fallback function to disallow any other deposits to the contract
        revert();
    }
    function Initialize(address payable _provider, address payable _consumer, uint _feePercent, 
    uint256 _blockNum) public payable onlyAuthorityNode  {
        require((_provider != msg.sender) && (_consumer != msg.sender));
        provider = _provider;
        consumer = _consumer;
        feePercent = _feePercent;
        blockNumber = _blockNum;
        EscrowStatus = EscrowState.initialized;

        EscrowAccountLedger[provider] = 0;
        EscrowAccountLedger[consumer] = 0;
    }
    function DepositInEscrowByConsumer() public payable checkBlockNumber onlyConsumer {
        EscrowAccountLedger[consumer] = SafeMath.add(EscrowAccountLedger[consumer], msg.value);
        deposits.push(msg.value);
        escrowCharge += msg.value;
        EscrowStatus = EscrowState.consumerDeposited;
        emit Deposit(msg.sender, msg.value); // solhint-disable-line
    }
    function ApproveEscrow() public {
        if (msg.sender == provider) {
            providerApproval = true;
        } else if (msg.sender == consumer) {
            consumerApproval = true;
        }
        if (providerApproval && consumerApproval) {
            EscrowStatus = EscrowState.serviceApproved;
            fee();
            EscrowPayout();
            emit ServicePayment(block.number, address(this).balance); // solhint-disable-line
        }
    }
    function CancelEscrow() public checkBlockNumber {
        if (msg.sender == provider) {
            providerCancel = true;
        } else if (msg.sender == consumer) {
            consumerCancel = true;
        }
        if (providerCancel && consumerCancel) {
            EscrowStatus = EscrowState.escrowCancelled;
            refund();
        }
    }
    function EndEscrow() public ifApprovedOrCancelled onlyAuthorityNode {
        DestructEscrow();
    }
    function DestructEscrow() internal {
        selfdestruct(authorityNode);
    }
    function EscrowPayout() private {
        EscrowAccountLedger[consumer] = SafeMath.sub(EscrowAccountLedger[consumer],
        address(this).balance);
        EscrowAccountLedger[provider] = SafeMath.add(EscrowAccountLedger[provider],
        address(this).balance);
        EscrowStatus = EscrowState.escrowComplete;
        providerAmount = address(this).balance;
        provider.transfer(address(this).balance);
    }
    function fee() private {
        uint totalFee = address(this).balance * (feePercent / 100);
        feeAmount = totalFee;
        authorityNode.transfer(totalFee);
    }
    function refund() private {
        consumer.transfer(address(this).balance);
    }
}

\end{lstlisting}
\twocolumn

\ifCLASSOPTIONcaptionsoff
  \newpage
\fi

\bibliographystyle{IEEEtran}
\bibliography{ref.bib}

\begin{IEEEbiography}[{\includegraphics[width=1in,height=1.25in,clip,keepaspectratio]{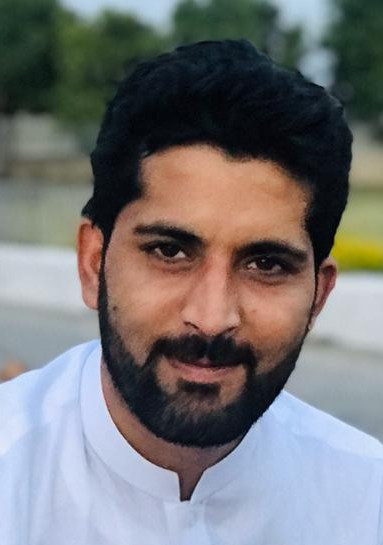}}]{Abdullah Yousafzai} is a postdoctoral research fellow under the prestigious grant of Brain Korea 21st Century Plus at the department of computer science and engineering, Kyung Hee University, Republic of Korea. Previously, he served as an assistant professor with the department of computer science and engineering, HITEC University, Taxila, Pakistan. Before that, he worked as a brightspark's research assistant at C4MCCR University of Malaysia, and as a backend web developer in Pakistan. He received his Ph.D. from the University of Malaya in 2017, MS (Computer Science) from Comsats Institute of Information Technology, Abbottabad in 2013 and BCS(Hons) from Hazara University Mansehra, Pakistan in 2009. His work mainly focuses on distributed computing environments comprising cloud computing systems, edge computing, mobile cloud computing, blockchain systems, and the Internet of Things.
\end{IEEEbiography}

\begin{IEEEbiography}[{\includegraphics[width=1in,height=1.25in,clip,keepaspectratio]{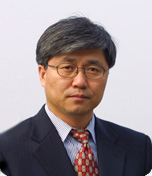}}]{Choong Seon Hong [S’95, M’97, SM’11]}  is working as a professor with the Department of Computer Science and Engineering, Kyung Hee University. His research interests include future Internet, ad-hoc networks, network management, and network security. He is a member of ACM, IEICE, IPSJ, KIISE, KICS, KIPS, and OSIA. He has served as the General Chair, a TPC Chair/Member, or an Organizing Committee Member for international conferences such as NOMS, IM, APNOMS, E2EMON, CCNC, ADSN, ICPP, DIM, WISA, BcN, TINA, SAINT, and ICOIN. In addition, he is currently an associate editor of the IEEE Transactions on Network and Service Management, the International Journal of Network Management, and the Journal of Communications and Networks and an associate technical editor of IEEE Communications Magazine.
\end{IEEEbiography}

\end{document}